\documentclass[11pt]{article}
\usepackage[letterpaper,hmargin=1in,vmargin=1.25in]{geometry}
\usepackage{times}
\usepackage{multirow}
\usepackage{graphicx}
\usepackage{amsmath,amsthm,amssymb,amsfonts,boxedminipage}
\makeatletter
\thm@headfont{\sc}
\makeatother
\newtheorem{athm}{\bf \t}

\begin{document}

\title{Enhanced No-Go Theorem for Quantum Position Verification}

\author{Fei Gao\footnote{gaofei\_bupt@hotmail.com}, Bin Liu, and Qiao-Yan Wen\\
State Key Laboratory of Networking and Switching Technology,\\ Beijing University of Posts and Telecommunications, \\
Beijing, 100876, China.\\
}

\maketitle
\thispagestyle{empty}

\begin{abstract}
Based on the instantaneous nonlocal quantum computation (INQC), Buhrman et al. proposed an excellent attack strategy to quantum position verification (QPV) protocols in 2011, and showed that, if the colluding adversaries are allowed to previously share unlimited entangled states, it is impossible to design an unconditionally secure QPV protocol in the previous model. Here, trying to overcome this no-go theorem, we find some assumptions in the INQC attack, which are implicit but essential for the success of this attack, and present three different QPV protocols where these assumptions are not satisfied. We show that for the \emph{general} adversaries, who execute the attack operations at every common time slot or the time when they detect the arrival of the challenge signals from the verifiers, secure QPV is achievable. This implies \emph{practically} secure QPV can be obtained even if the adversaries is allowed to share unlimited entanglement previously. Here by ``practically'' we mean that in a successful attack the adversaries need launch a new round of attack on the coming qubits with extremely high frequency so that none of the possible qubits, which may be sent at random time, will be missed. On the other side, using such Superdense INQC (SINQC) attack, the adversaries can still attack the proposed protocols successfully in theory. The particular attack strategies to our protocols are presented respectively. On this basis, we demonstrate the impossibility of secure QPV with looser assumptions, i.e. the enhanced no-go theorem for QPV. 
\end{abstract}
\section{Introductions}

\subsection{Background}
As an interesting branch of cryptography, position-based cryptography (PBC) \cite{PBC} adds an additional layer of security in the sense that the geographical position of a party is used as its credential. That is, it guarantee that only the users in an intended geographical position can achieve the current cryptographic goals. PBC might have potential applications in military communications and our daily life.

Position verification (PV) is one of the important applications of PBC. It allows a prover to prove to a set of cooperating verifiers, who are spatially separated, that he/she is at a particular spatial location. Because of its great usefulness in some interesting applications, PV has drawn a lot of attention and quite a few schemes were proposed for that (see Ref. \cite{PBC} for details). However, Chandran et al. proved that secure PV protocol are impossible in the classical world \cite{PBC}. In view of the huge success of quantum cryptography on its security, it is desirable to introduce quantum mechanics into PV to resolve the above problem.

Quantum position verification (QPV) was first studied by Kent under the name of ``quantum tagging'' as early as 2002. In 2006 a patent of quantum tagging given by Kent et al. was granted \cite{Pat}. But their idea was not publicly announced as a paper until 2010 \cite{KMS10}. Just before the appearance of Ref. \cite{KMS10}, Chandran et al. \cite{CFG10} and Malaney \cite{M101} independently proposed a QPV protocol, respectively. In Ref. \cite{KMS10} Kent et al. pointed out the insecurity of some primary QPV protocols, including the ones in Refs. \cite{CFG10,M101}, and then presented two improved protocols which can stand against their attacks. Lau and Lo studied the security of the protocols in Refs. \cite{CFG10,M101} further and gave general attack strategies to them in detail. Furthermore, they proposed a modified scheme and proved its security under the condition that the shared quantum resource between the adversaries is a two- or three-level system \cite{LL11}. However, utilizing the technology of instantaneous nonlocal quantum computation (INQC), Buhrman et al. recently proposed the no-go theorem for QPV, where all the previous QPV protocols were proved insecure under the condition that unlimited entanglement were previously shared between the adversaries \cite{Buhrman,Nature}. Afterwards, the INQC attack was improved in the sense that the amount of the required entanglement for a successful attack is reduced \cite{BK11}. In addition, Buhrman et al. defined a new model of communication complexity, the garden-hose model, which can be used to prove upper bounds on the number of EPR pairs needed to attack QPV protocols \cite{gardenhose}.

\subsection{The previous models of QPV protocols}

\begin{figure}
  \center
  \includegraphics[width=14cm]{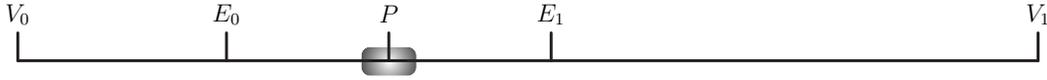}\\
  \caption{One-dimension QPV and the INQC attack. The shaded area is the finite secure region around $P$.}\label{QPV-INQC}
\end{figure}

In a QPV protocol a prover $P$ wants to convince a set of $k$ verifiers $V_0, \ldots, V_{k-1}$, who are located at different reference stations, that he/she is at a particular spatial location. In this condition the aim of the adversaries is to persuade the verifiers that $P$ is working in the correct position while $P$ has actually been removed. The setting of the simplest QPV is the one-dimension one, where two verifiers $V_0$ and $V_1$ want to verify the position of the prover $P$ (all of them are in a straight line and $P$ is located between $V_0$ and $V_1$, see Fig.~\ref{QPV-INQC}).

Fig. \ref{Model} describes a general model of QPV, which we call standard QPV (SQPV). Till now, most protocols were designed in this model \cite{KMS10,CFG10,M101,LL11}. The only exception is the recent one give by Kent \cite{Kent11}, where the item R7 is removed. That is, the prover $P$ has a secret key which helps he/she to show his/her unique identity to the verifiers. This model, which we call keyed QPV (KQPV), is actually a compromise between the requirements in real applications and the security against the INQC attack (generally, $P$ is only a device and might be obtained by the adversaries, so it is difficult to guarantee the security of the key in it). Therefore, generally speaking, SQPV is more desirable if it can keep secure under all potential attacks.
\begin{figure}
\begin{boxedminipage}{1\textwidth}
\begin{itemize}
  \item[R1] All the signals including the quantum and classical ones are transmitted at light speed.\vspace{-2.5mm}
  \item[R2] The verifiers have synchronized clocks.\vspace{-2.5mm}
  \item[R3] The sizes of all the users' devices are negligible.\vspace{-2.5mm}
  \item[R4] $P$ has a finite secure region around him, where the adversaries have no access. (Otherwise, any one can ``personate'' $P$ by just standing at $P$'s position.)\vspace{-2.5mm}
  \item[R5] All the computations, including classical computations, quantum measurements, and quantum operations, are instantaneous. That is, the times consumed by all these computations are negligible.\vspace{-2.5mm}
  \item[R6] To authenticate the prover's position, the verifiers can cooperate with each other before and after the communication with the prover. For example, they can previously share some secret information, and perform some calculations or measurements together after the communication to determine the verification result.\vspace{-2.5mm}
  \item[R7] There is no secret which is unknown for the adversaries between the verifiers and $P$.
\end{itemize}
\end{boxedminipage}
\caption{The standard QPV model.}\label{Model}
\end{figure}

The QPV protocols work in a challenge-response manner. That is, the verifiers send a challenge command to the prover and judge the correctness and the timeliness of his/her corresponding response. The command might be to measure a given quantum system in a certain basis and announce the result, or to perform a certain operation on the quantum system and send it back to the verifiers. The command is generally divided into several pieces and sent by different verifiers respectively. The sending times of all these pieces are controlled so that they will arrive at the intended position simultaneously. Thus the person indeed in this position is the first one who can obtain all the pieces, recover the complete command, and then give the correct response in time. While the one in other positions can only recover the command and give a correct response in a later time (note that if he/she tries to make the response arrive at the verifiers in time its correctness will not be assured because he/she has to send it out before all the pieces are obtained).

Generally, the challenge-response process is executed round by round. If all the responses are right and in time in enough rounds, the verifiers will accept the prover's position. Concretely, as defined in Ref. \cite{Buhrman}, the process of each round in QPV protocol can be described as in Fig. \ref{SM}.

\begin{figure}
\begin{boxedminipage}{1\textwidth}
For each round, there are four quantum systems $A$, $B$, $J$, and $R$. $P$ holds the system $R$, which is set to some default value at the beginning. $V_0$ keeps $A$, $V_1$ keeps $B$, and they jointly keep $J$. There are also two classical information $x$ and $y$, belonging to $V_0$ and $V_1$ respectively. As we will see, $A$, $B$, $x$, and $y$ are the challenge signals, i.e. the pieces of the challenge command, which will be sent to $P$ so that he/she can recover the complete command from them.\vspace{-2mm}
\begin{itemize}
  \item[G1] $V_0$ sends $A$ and $x$ to $P$, and $V_1$ sends $B$ and $y$ to $P$. The sending time is controlled so that $P$, the one who is really in the intended position, will receive all of them simultaneously.\vspace{-2mm}
  \item[G2] According to the classical information $x$ and $y$, $P$ performs different operations or measurements to $ABR$ as soon as he/she received these challenge signals, and then sends the response (e.g. the resulted systems or the measurement result) to $V_0$ and $V_1$, respectively.\vspace{-2mm}
  \item[G3] $V_0$ and $V_1$ jointly judge whether the response they received is correct and timely.
\end{itemize}
\end{boxedminipage}
\caption{One round of the one-dimension standard QPV.}\label{SM}
\end{figure}

\subsection{The INQC attack to QPV}

In Buhrman et al.'s INQC attack \cite{Buhrman}, Two adversaries $E_0$ and $E_1$ stand in the line between $V_0$ and $P$, and $V_1$ and $P$ respectively, and, for simplicity, the distances between $E_j$ ($j=0,1$) and $P$ are equal (see Fig. \ref{QPV-INQC}). In every challenge-response round, $E_0$ and $E_1$ can simultaneously intercept the challenge signals $(A, x)$, sent by $V_0$, and $(B, y)$, sent by $V_1$, respectively. To successfully cheat the verifiers, they have to perform nonlocal measurement or operation on $A$ and $B$ in the condition that only one mutual communication can be done between them. The INQC attack, in which $E_0$ and $E_1$, using the pre-shared entanglement between them, perform teleportations (the classical communication in the general teleportation is not included here) round by round, can perfectly achieve the above task the adversaries face to (see Sec.2.3 for details). Therefore, Buhrman et al. drew the conclusion that designing QPV protocol with unconditional security is impossible in the standard security model, where the pre-shared entanglements between the adversaries are unlimited.

Because of its significance, the no-go theorem has great influence in this field. To overcome it, Buhrman et al. gave another security model instead of the standard one, i.e. the bounded model, which only allows limited pre-shared entanglements between the adversaries \cite{Buhrman}. Recently, Kent pointed out that secure QPV can be still achieved if the prover previously shared a secret key with the verifiers \cite{Kent11}. To some extent, however, the requirement of a keyed prover seems too strong. Can we find a new way to overcome the no-go theorem without the above restrictions? This is a very important problem which helps us to know what kind of security the quantum mechanics can really bring us.

\subsection{Our results}
Though giving a secure SQPV in the bounded model or a secure KQPV protocol is possible, the most challengeable thing is still to design a secure SQPV in the standard secure model. The problem is whether we can find some special manners to get away from the application of the INQC attack. An effective approach to this problem is to find out the hidden assumptions in the INQC attack, which is essential for a successful attack, and then explore new QPV protocols which are not consistent with these assumptions. Based on this idea, we obtain the following results in this paper.

\textbf{The Hidden Assumptions.}
By careful study on the INQC attack, we find four crucial assumptions in it. Breaking those assumptions in designing a QPV protocol might be an effective way to overcome the INQC attack (see Sec.3).

\textbf{The Proposed QPV Protocols.} We propose three types of new QPV protocols, which are not consistent with the above hidden assumptions (see Secs.4-6). Our analysis shows that for the \emph{general} adversaries, who execute the attack operations at every common time slot or the time when they detect the arrival of the challenge signals from the verifiers, all the proposed protocols are secure even if the adversaries are allowed to share unlimited entanglement previously.

\textbf{The Superdense INQC attack.} Though it is very difficult to implement in practice, the adversaries can adopt an improved INQC attack, i.e. the superdense INQC (SINQC) one, to successfully attack our QPV protocols in theory. In SINQC attack, the adversaries need launch a new round of attack on the coming qubits with extremely high frequency so that none of the possible qubits, which may be sent at random time, will be missed. We present the particular SINQC attacks for our protocols, respectively, which are quite different with previous INQC attack (see Secs.4-6).

\textbf{The Enhanced No-Go Theorem.} Based on the success of the SINQC attack to our protocols, we draw a conclusion that secure QPV is still impossible under looser assumptions in theory, which we call the the enhanced no-go theorem for QPV (see Sec.7).

It should be emphasized that, though our original aim, i.e. overcoming the no-go theorem in QPV, has not been achieved at last, the enhanced no-go theorem we obtained, including the ideas of our protocols and corresponding SINQC attacks, are new and interesting. We hope that our results will shed light in the future study of quantum position-based cryptography.

\section{Preliminaries}
\subsection{Mach-Zehnder Interferometer and Single-photon Interference}
\begin{figure}
  \center
  \includegraphics[width=6cm]{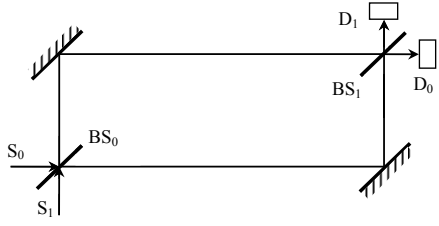}\\
  \caption{Single-photon interference in a Mach-Zehnder Interferometer. BS$_0$ and BS$_1$ are 50:50 Beam Splitters.}\label{fig2}
\end{figure}

Single-photon interference is an interesting phenomenon in the quantum world. Fig. \ref{fig2} is the demonstration of a Mach-Zehnder Interferometer. When a photon is sent to BS$_0$ from the source S$_0$, two localized wave packets appear in the upper and the lower channels respectively. The state of the quantum system becomes $|\Phi_0\rangle=\frac{1}{\sqrt{2}}(|10\rangle+|10\rangle)$ where the two qubits represent two channels respectively, and 0 means a vacuum state and 1 implies a photon. When the two wave packets arrive at BS$_1$ simultaneously, interference will happen and the photon will be detected by the detector D$_0$ with certainty. Similarly, if the photon comes from S$_1$, the state after BS$_0$ will be $|\Phi_1\rangle=\frac{1}{\sqrt{2}}(|10\rangle-|10\rangle)$ and the photon will be detected by D$_1$ after the interference at BS$_1$. In this process, if one performs measurement in the channels to know which way  the photon is going through, the interference will not happen any more. That is, the which-way information and the relative phase between the two wave packets can never be (exactly) obtained simultaneously. This is an fundamental principle in quantum mechanics, i.e. complementarity \cite{Complem}. Obviously this principle put a special limit on the states $|\Phi_0\rangle$ and $|\Phi_1\rangle$, making them quite different from general two-particle Bell states. In fact, the application of this peculiarity in cryptography has been noticed for a long time. For, example, Goldenberg and Vaidman used it to design a novel quantum key distribution protocol in 1995 \cite{GV95}.

\subsection{Port-based teleportation}
Port-based teleportation (PBT) was introduced by Ishizaka and Hiroshima \cite{PBT1,PBT2}, and then was used to reduce the entanglement resource required in the INQC attack by Beigi and K\"{o}nig \cite{BK11}. Different from the better known teleportation scheme where Bob performs certain Pauli operations to do the correction according to Alice's classical information (i.e. the Bell-measurement result), PBT employs a simpler correction on Bob's side. That is, Bob just discard some subsystems according to Alice's classical information.

Suppose that Alice wants to teleport a state $|\Psi\rangle_A\in\mathbb{C}^d$ to Bob. In advance, Alice and Bob share a resource state, which consists of $N$ maximally entangled pairs in the state $|\Phi\rangle_{A^\prime_iB^\prime_i}=\frac{1}{\sqrt{d}}\sum^d_{i=1}|i\rangle|i\rangle$, where $i=1,2,\ldots,N$.
Then PBT can be done as follows.
\begin{itemize}
  \item[T1] Alice performs a certain POVM $\{E^i\}^N_{i=1}$ on her systems $AA^\prime_1A^\prime_2\ldots A^\prime_N\in (\mathbb{C}^d)^{\otimes {N+1}}$. And she obtains the measurement outcome $k$ ($1\leq k\leq N$) and sends it to Bob.\vspace{-2mm}
  \item[T2] Bob discards all the port subsystems $B^\prime_i (i\neq k)$ except for $B^\prime_k$. And $B^\prime_k$ now contains the teleported state $|\Psi\rangle$.
\end{itemize}
Note that this kind of teleportation does not inevitably succeed, and the success probability increases as $N$ gets larger (close to 1 for large $N$).

\subsection{INQC attack utilizing PBT}
Because the PBT-based description of the INQC attack looks more understandable we will describe the attack strategies to QPV in such a way throughout the paper.
Note that in our description both $E_0$ and $E_1$ use PBT, which is different from that in Ref. \cite{BK11}, where $E_0$ uses usual teleportation and $E_1$ uses PBT.

Suppose there are totally $m$ rounds in the QPV protocol, and each round proceeds as in Fig.~\ref{SM}.
For simplicity, we suppose the challenge signals are all quantum (for the classical parts $x$ and $y$, the adversaries can encode them into quantum ones in the computational basis), and the distance between $E_0$ and $P$ is equal to that between $E_1$ and $P$. The notations we will use in the description are listed in Table \ref{tab1}. In the INQC attack \cite{Buhrman}, the adversaries perform the following operations in every round. For example, in the $i$-th round,

\begin{table}
\centering
\caption{The notations in the INQC attack. Here $j,k=0,1$ and $i=1, 2, \ldots, m$.}
\label{tab1}\vspace{1mm}
\renewcommand{\arraystretch}{1.15}
\begin{tabular}{c|l}
  Notation & What the notation denotes \\
  \hline\hline
  \multirow{1}*{$Q_j^i$} & The challenge signal received by $E_j$ in the $i$-th round (i.e. $A$ and $x$, or $B$ and $y$ in Fig.~\ref{SM}).\\\hline
  \multirow{1}*{$\rho_j^i$} & The quantum state of $Q_j^i$.\\\hline
  \multirow{1}*{$B^i$} & The pre-shared entangled systems by the adversaries used to perform PBT in the $i$-th round.\\\hline
  \multirow{1}*{$B^{i,k}$} & The $k$-th subgroup of $B^i$ used for the $k$-th PBT in the $i$-th round. \\\hline
  \multirow{1}*{$B_j^{i,k}$} & The subsystems of $B^{i,k}$ on $E_j$'s side.\\\hline
  \multirow{1}*{$n_j^i$} & The which-port-information of the PBT $E_j$ performed in the $i$-th round.\\ \hline
  \multirow{2}*{$\rho(n_j^i)$} & The state of the resulted multi-port system after the counterpart's measurement in PBT. The\\
                               & dimension of every port is equal to that of $\rho$ and the state of the $n_j^i$-th port is just $\rho$.\\\hline

\end{tabular}
\end{table}

\begin{itemize}
\item[$S_1$] When $Q_0^i$ and $Q_1^i$ simultaneously arrive at $E_0$ and $E_1$, respectively, $E_0$ performs a measurement on the systems $Q_0^iB_0^{i,1}$ to execute a PBT of $Q_0^i$ with the outcome $n_0^i$. And then $E_1$ holds $\rho_0^i$ at the $n_0^i$-th port in her system $B_1^{i,1}$ [the state of the whole $B_1^{i,1}$ now is $\rho_0^i(n_0^i)$].\vspace{-2mm}
\item[$S_2$] $E_1$ performs a measurement on the systems $B_1^{i,1}Q_1^iB_1^{i,2}$ to teleport $B_1^{i,1}Q_1^i$, which is in the state $\rho_0^i(n_0^i)\rho_1^i$, with the outcome $n_1^i$. And then $E_0$ holds
    \begin{equation}\label{e3}
        [\rho_0^i(n_0^i)\rho_1^i](n_1^i)
    \end{equation}
    in $B_0^{i,2}$.\vspace{-2mm}
\item[$S_3$] For each port in $B_0^{i,2}$, $E_0$ discards everything except the $n_0^i$-th sub-port to do the correction to the chaos introduced by herself in the PBT she did above, obtaining the state
    \begin{equation}\label{e3-1}
        [\rho_0^i\rho_1^i](n_1^i)
    \end{equation}
    in the reminder system. Note that now $\rho_0^i$ and $\rho_1^i$ have been put together in the $n_1^i$-th port at $E_0$ (but $E_0$ does not know $n_1^i$ till now).\vspace{-2mm}
\item[$S_4$] For every port in the reminder system, $E_0$ reads the classical signals (i.e. $x$ and $y$) by measurements on the corresponding subsystems, and then executes the operations according to the command (typically, performs measurement or unitary operations on the quantum signals according to the classical ones). Afterwards $E_0$ sends the measurement result, or the resulted qubits (if these qubits should be sent to $V_1$ according to the command) to $E_1$.
\end{itemize}
\hspace{0.5cm}

Note that all the above operations by $E_0$ and $E_1$ are executed simultaneously, and they do not consume time. At the same time, the adversaries also send the received classical signals (i.e. $x$ and $y$) and the which-port-information in their respective PBT (i.e. $n_0^i$ and $n_1^i$) to each other. Once they received all these contents, both $E_0$ and $E_1$ know the correct response to the command. Till now, the difficulty the adversaries faced to in the attack, that is, nonlocal measurement or operation on the separated quantum signals in the condition that only one mutual communication can be done between them, has been overcome. Finally $E_0$ and $E_1$ then send their responses to $V_0$ and $V_1$ respectively. Obviously the response will be correct and timely, and the attack will be successful. We emphasize that the above description of the INQC attack is a general version, where some transmissions (e.g. $x$, $y$, and $n_0^i$) or teleportations (e.g. $x$) are actually unnecessary for the QPV protocols considered here.

\section{The hidden assumptions in the INQC attack}
By careful observation We find the following four hidden assumptions in the INQC attack.
\begin{itemize}
  \item[H1] The quantum parts of the challenge signals consist of normal qubits in the previous QPV protocols, which can be teleported by the adversaries without any disturbance on their states.\vspace{-1mm}

      \textbf{Analysis}. If the verifiers use wave packets divided from one photon instead of normal qubits as the quantum signals, and require the prover to observe their interference (like GV95 protocol \cite{GV95}), what will happen? Can the wave packets be teleported by the adversaries? If it can be done, whether the measurements in teleportation, which generally means the discovery about which way the photon is passing through, would disturb the interference? Note that the complementarity principle forbids simultaneously observing the which-way information and the interference in this situation \cite{Complem}.\vspace{-2mm}
  \item[H2] The adversaries are aware of when to launch a new round of attack operations in the INQC attack.\vspace{-1mm}

      \textbf{Analysis}. Generally speaking, the adversaries launch a new round of attack operations when a new pair of challenge signals is coming. If the challenge signals are sent at fixed time slots, the adversaries can perform the teleportation at the fixed time points when the challenge signals would arrive at their positions. Alternatively, if normal qubits are employed, the adversaries can begin their teleportation of the challenge signals when they detect the qubits. However, if the verifiers send the challenge signals at random time instead of the fixed time slots, and use wave packets instead of normal qubits in them (so that the adversaries cannot know their coming because discovering the coming of any wave packet means the detection of the photon, which will inevitably disturb the interference), both the above conditions are unsatisfied. Then what will happen?\vspace{-2mm}
  \item[H3] The different rounds of the protocol are independent with each other.\vspace{-1mm}

      \textbf{Analysis}. QPV protocol can be designed so that the prover's result $R_p$ in the previous round is also a piece of command in the present round. But it is trivial to break down this assumption \emph{separately} because it is just equivalent to that when combining the challenge signals in the adjacent two rounds as the command of the present round. In fact, Buhrman et al. also considered this situation in Ref. \cite{Buhrman}. However, as we will show in Secs.5-6, combining it with other manners (i.e. breaking down other assumptions) is a possible way to overcome the INQC attack. So we still list this assumption here.\vspace{-2mm}
  \item[H4] The adversaries are aware of which challenge signals are matched signals. Here ``matched'' signals means the ones that compose a complete command and will be operated collectively by the prover. For example, in Fig.~\ref{SM} $(A,x)$ and $(B,y)$ are a pair of matched signals.\vspace{-1mm}

      \textbf{Analysis}. As we know, both the QPV protocol and the INQC attack are executed round by round, and the separate adversaries need to use the same group of the shared entangled states to deal with the matched signals [for example, in the INQC attack in Sec.2.3, both $E_0$ and $E_1$ use $B^i$ to teleport their respective (matched) signals $Q^i_0$ and $Q^i_1$]. Therefore, if the adversaries do not know which signals are matched ones, they might use different group of entangled states to deal with a pair of matched signals, which obviously will result in the failure of the INQC attack.

      In fact, recognizing the matched signals is a quite simple task so that it is always overlooked in the study. Generally, as in all the previous QPV protocols, the matched signals will be sent in fixed time slots by $V_0$ and $V_1$ and reach $P$ simultaneously. In this condition $E_0$ and $E_1$ knows that the signals they received in the same time slot are matched ones. Even though the signals are sent at random time, as discussed in H2, $E_0$ and $E_1$ still knows that the signals they received in the same order are matched ones. However, if we make some of the matched signals encounter at other positions (e.g. $P'$) instead of $P$ (note that it does not damage the security of QPV when $P'$ is also in $P$'s secure region), the signals received in the same time slot by $E_0$ and $E_1$ might not matched ones anymore. Alternatively, $V_0$ and $V_1$ can also insert some decoy signals in the ordinary ones so that the signals received by $E_0$ and $E_1$ in the same order are not matched. So, if we do such modifications in the QPV protocols, can INQC attack still work?
\end{itemize}

As is analyzed above, the key to overcome the no-go theorem for QPV is to break down these hidden assumptions. Based on this idea, we will present QPV protocols which are not consistent with them, and then analyze their security in detail.

\section{Protocol I: Attempts to break down H1 and H2}
\begin{figure}
\center
  \includegraphics[width=7cm]{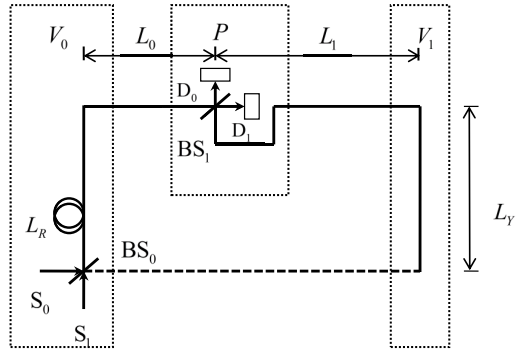}\\
  \caption{Protocol I. The dotted rectangle implies the user's lab, where no adversaries can enter. The circles in $V_0$'s lab are the storage rings with length $L_R$. The interferometer is a balanced one, that is, the lengths satisfy $L_Y+L_R+L_0=L_0+L_1+L_Y+L_1$.}\label{PI}
\end{figure}

Based on the analyses about H1 and H2, we present a QPV protocol (called Protocol I) based on single-photon interference and random sending time, which breaks down the two assumptions simultaneously. Protocol I proceeds as follows (See Fig. \ref{PI}).

\begin{itemize}
  \item[I1] $V_0$ sends single photons into the interferometer, at random time, from $S_0$ or $S_1$ with equal probability. $V_0$ records the sending time of every photon.\vspace{-2mm}
  \item[I2] To ensure the security of the transmission in the lower channel, $V_0$ and $V_1$ can measure the wave-packet pairs, which are generated at BS$_0$ in some random (can also be predetermined) time periods, and judge whether there are one and only one photon for each pair.

      \textbf{Note}. Generally, how to secretly share the wave-packet pairs between $V_0$ and $V_1$ is not the focus in the QPV protocol, and consequently Steps I1 and I2 can be omitted here (see R6 in Fig.~\ref{Model}). However, to represent the protocol clearly, we still list them here and in Fig. \ref{PI}.\vspace{-2mm}
  \item[I3] The other wave-packet pairs (which are not measured by $V_0$ and $V_1$) will appear at $P$'s detectors $D_0$ or $D_1$ via interference. Once $P$ received a photon, he broadcasts the value (i.e. which detector the photon appear at) to $V_0$ and $V_1$ right now.\vspace{-2mm}
  \item[I4] $V_0$ and $V_1$ record the receiving time of $P$'s response and the value in it. After sending enough photons, $V_0$ and $V_1$ authenticate the position of $P$ according to the receiving times and the values. If the responses for all photons are in time and correct (and no eavesdropping is detected in step I2), they believe $P$'s position. Otherwise, the position does not pass the authentication.
\end{itemize}

Now we consider the possible problems, as analyzed in H1 and H2, that might appear in the INQC attack to such a QPV protocol.

\textbf{Problems in H1.} Can the wave packets be teleported? If it can be done, whether the measurements in the adversaries' teleportation, which generally means the discovery about which way the photon is passing through (that is, the measurement result of any adversary inevitably implies the number of the coming photons, i.e. 0 or 1), would disturb the interference, and consequently they can never obtain the correct response?

Successful teleportation is an essential requirement of the INQC attack. Because there might be no photon in a wave packet at all, can it be teleported? If the answer is negative, the INQC attack will be useless to Protocol I. However, in fact, the adversaries can change the wave-packet state $|\Phi_0\rangle$ or $|\Phi_1\rangle$ into a general two-qubit entangled state in theory \cite{Aharonov}, and consequently the state can be teleported and such teleportation need not reveal the which-way information about the photon. Therefore, it is unnecessary to worry about the above two problems when the adversaries execute the INQC attack to Protocol I.

\textbf{Problem in H2.} Can the adversaries be aware of when to launch a new round of attack operations in the INQC attack?

Obviously the using of single-photon interference and random sending time will bring new difficulty for the adversaries in INQC attack. On the one hand, the challenge signals (wave packets) are sent at random time points instead of fixed time slots, the adversaries cannot perform the teleportation at the fixed time points when the challenge signals would arrive at their positions. On the other hand, the way that using the detection of a signal's coming to give rise to a new round of attack is not feasible, either. If the adversaries want to judge whether a photon is coming, according to the complementarity principle, their observation will expose the which-way information and then result in that no interference will happen later. (In fact, there is only one photon for each pair of wave packets in the two channels, so it is impossible for any pair to tell both $E_0$ and $E_1$ to begin a new attack round.) As a result, in the INQC attack to Protocol I, the adversaries will not recognize when to launch a new round of attack operations as before.

It should be emphasized that the prover $P$ need not know the time when the pair of wave packets will arrive at his/her position. This is because the two packets will reach $P$ simultaneously and then be detected by $D_0$ or $D_1$. That is, $D_0$ or $D_1$ will be triggered automatically once the packets arrived. But for the adversaries $E_0$ and $E_1$, things are different. They have to know the coming of a wave packet so that they can launch the teleportation. One may ask a question as follows. Is it possible that the adversaries put a device (including qubits from the shared entangled states), waiting the coming of the signal to do the measurement for teleportation of it also in an automatical way? The answer is negative. Obviously, there are the following two conditions about this measurement.\vspace{-2mm}
\begin{itemize}
  \item The measurement is triggered by the coming photon. That is, if there is no signal coming, the measurement will not be performed. In this condition, once the measurement device clicked, $E_1$ knows that there is a photon, that is, she knows the which-way information, so, she will not obtain the correct relative phase and consequently their attack will fail.\vspace{-2mm}
  \item The measurement is NOT triggered by the coming photon. Then, to catch every signal passing through, $E_0$ and $E_1$ have to perform the attack continually for any time (no matter whether there is a signal in the channel at that time).
\end{itemize}

Now we can draw a conclusion about the security of Protocol I. On the positive side, it is secure against INQC attack for the \emph{general} adversaries, who execute the attack operations at every common time slot or the time when they detect the arrival of the challenge signals from the verifiers. This implies \emph{practically} secure QPV can be obtained even if the adversaries is allowed to share unlimited entanglement previously. Here by ``practically'' we mean that an successful attack exists only in theory because the adversaries have to perform superdense INQC (SINQC), that is, to launch a new round of attack on the coming qubits with extremely high frequency so that none of the possible qubits, which may be sent at random time, will be missed. As a interesting result, \emph{by using single-photon interference and random sending time (i.e. breaking down H1 and H2), the adversaries in INQC attack are forced to be blind}. This additional difficulty for a successful attack just like that a blind man, who stands beside a fiber where photons will pass by at random time, tries to ``cut'' the fiber by a knife continually so that he can ``catch'' all the possible photons.

On the other side, though it is very difficult to achieve in practice the adversaries can still attack the proposed protocol successfully by SINQC. So Protocol I is still insecure in theory. Nevertheless, except for the difficulty discussed above, the adversaries need much more shared entanglement in such attack because most consumed entangled states are wasted (used on the signals which do not exist at all).

\section{Protocol II: Attempts to break down H3 and H4}

In this section, we mainly consider to shatter the hidden assumptions H4, and when needed, the breach of H3 will be also considered together. As discussed in H4, one prerequisite for the INQC attack is that the adversaries are aware of which pair of signals are matched. If the matched challenge signals are arranged that they arrive at $P$ at different times, the separated adversaries will not know which signals are matched ones, and might use different group of entangled states to deal with a pair of matched signals, which obviously will result in the failure of the INQC attack. For simplicity, we call this kind of protocols Different-Time QPV (DTQPV), which is quite different from the previous protocols. Here we should emphasize that, though in DTQPV the matched signals are not simultaneously arrive at the position of $P$ any more, the protocol can be still secure like that in the previous QPV protocols. For example, it can be designed that to receive the complete matched signals the person at any position out of the secure region of $P$ will be later than $P$).

As we know, if the every pair of matched signals are sent from $V_0$ and $V_1$ independently in turn, $E_0$ and $E_1$ can still execute INQC attack according to the order of the signals she received even though the two matched signals arrive at $P$ at different times. That is, as discussed in H2 and H4, $E_0$ and $E_1$ still knows that the signals they received in the same order are matched ones. Therefore, there generally are some additional manners to protect the protocol against the above attack by $E_0$ and $E_1$ in DTQPV. For example, $V_0$ and $V_1$ can also insert some decoy signals in the ordinary ones so that the signals received by $E_0$ and $E_1$ in the same order are not matched. We need not worry about how $P$ distinguishes the decoy signals and discards them because this information can be obtained from the measurement result on the previous matched signals [e.g. the measurement result 0 (1) means the next signal from $V_0$ ($V_1$) is a decoy one and should be overlooked]. Obviously in this condition the assumption H3 is also broken down.

In the following we will present two kinds of DTQPV protocols and discuss their security in detail.

\subsection{The public-order DTQPV protocol}

Though DTQPV looks like an effective model to stand against the INQC attack, designing secure DTQPV protocols (with normal qubits) is challenging. Here we introduce a special kind of DTQPV protocol, i.e. the so-called public-order DTQPV, where the chronological order for all the challenge signals (from both sides) arriving at $P$ is known by both $P$ and the adversaries. As we will show, this kind of protocol are still insecure for general INQC attack. But we need talk about it (especially the corresponding attack) in detail because the attack strategy is a basis for the analysis of a general DTQPV which will be given in the next subsection.

Since the challenge signals reach $P$ in a public order, we can label all of them according to the chronological order of their arrivals at $P$ under the following rules. (1) If there come two signals from two sides of $P$ simultaneously, the challenge signal from $V_0$ should be labeled anterior to the one from $V_1$. (2) If there come two adjacent signals from one side, which means that no signal comes from the other side between them, a special signal from the other side, denoted as an ``empty signal'', is added factitiously between them (note that this is can be achieved because the adversaries is assumed to know the order of the challenge signals here). And the added empty signal is assumed to reach $P$ at the same time with the anterior one. (3) Assume that the first signal is from $V_0$, otherwise, add an empty signal from $V_0$ before the real first one. In this way, all the challenge signals can be listed according to the chronological order as follows.
\begin{equation}\label{e0}
Q_0^1Q_1^2Q_0^3Q_1^4\ldots Q_0^{n-1}Q_1^n
\end{equation}
Here, the subscript $0$ represents the challenge signal is from $V_0$, correspondingly $1$ represents the challenge signal is from $V_1$, and the superscripts represent the challenge signal's order. Obviously, following the three rules above, the challenge signals, including the empty ones, arrive at $P$ from $V_0$ and $V_1$ alternately one by one now. Here we do not limit that only the adjacent two signals can compose the matched ones. Maybe ($Q_0^1$, $Q_0^3$), ($Q_0^1$, $Q_1^4$), or even ($Q_0^1$, $Q_0^3$, $Q_1^4$) are matched signals as long as such a protocol is still secure like that in the previous QPV protocols (e.g, to receive the complete matched signals the person at any position out of the secure region of $P$ will be later than $P$). This can be taken as the general model of the public-order DTQPV. Note that the (quantum) signals involved here is normal qubits, which can be detected without disturbing their states.

Now we will introduce our proposed attack to this kind of protocols. For simplicity, we suppose the challenge signal $Q_j^i$ is all quantum and in the state $\rho^i$, where $j=0,1$ (for the classical parts, the adversaries can encode it into quantum ones). To describe it clearly, we first list the notations that we will use in the following description of the attack strategy in Table \ref{tab2}.
\begin{table}
\centering
\caption{The notations in the description of Attack I. Here, $j=0,1$ and $i=1,\ldots, n$.}
\label{tab2}\vspace{1mm}
\renewcommand{\arraystretch}{1.15}
\begin{tabular}{c|l}
  Notation & What the notation denotes \\
  \hline\hline
  \multirow{1}*{$B^i$} & The pre-shared entangled systems $E_j$ used to perform the teleportation when $Q_j^i$ arrives.\\\hline
  \multirow{1}*{$B_j^i$} & The subsystems of $B^i$ on $E_j$'s side.\\\hline
  \multirow{1}*{$n_j^i$} & The which-port-information of the teleportation $E_j$ performed when $Q_j^i$ arrives.\\\hline
  \multirow{2}*{$\rho(n_j^i)$} & The state of the resulted multi-port system after the counterpart's measurement in PBT. The\\
                             & dimension of every port is equal to that of $\rho$ and the state of the $n_j^i$-th port is just $\rho$.\\\hline
  \multirow{2}*{$G^i$}       & The operation that the adversaries should do on the quantum part according to the classical part\\
                             & in one port, when he receives $Q^i_j$, including measurements, unitary operations or nothing.\\\hline
  \multirow{2}*{$L^i$} & The information (for all the ports) that should be sent to the counterpart after $G^i$, e.g. the\\
                             & measurement results or the operated qubits which should be sent to the verifier at the other side. \\\hline
  \multirow{1}*{$R_j^i$} & The remainder system of the previous steps at $E_j$ when $Q_j^i$ arrives (see the attack for details). \\\hline
\end{tabular}
\end{table}
In this attack (Attack I), $E_j$ performs the following operation $A_i$ when the challenge signal $Q_j^i$ arrives.
\begin{itemize}
\item[$A_1$] When $Q_0^1$ arrives, $E_0$ performs a measurement on the systems $Q_0^1B_0^1$ to execute a PBT of $Q_0^1$ with the outcome $n_0^1$. And then $E_1$ holds $\rho^1$ at the $n_0^1$-th port in system $B_1^1$ [the state of the whole $B_1^1$ now is $\rho^1(n_0^1)$]. We suppose that $E_1$ has performed the next teleportation utilizing $B^2$ (this teleportation will be done when $Q_1^2$ comes to $E_1$, and it will be really executed in Step $A_2$). And $E_0$ now holds
    \begin{equation}\label{e1}
        [\rho^1(n_0^1)\rho^2](n_1^2)
    \end{equation}
    in $B_0^2$. The above supposition is reasonable since the measurement results to the local systems are independent with the measurement order. Simultaneously, for each port in $B_0^2$, $E_0$ first discards everything except the $n_0^1$-th sub-port to do the correction to the chaos introduced by herself and the state changes to
     \begin{equation}
        [\rho^1\rho^2](n_1^2).
     \end{equation}
     And then $E_0$ performs the operation $G^2$ (i.e. the operation or measurement she should do according to the classical part of the signals, nothing should be done if $\rho^1$ and $\rho^2$ are not matched) on all the ports respectively. Afterwards, $E_0$ sends the necessary information $L^2$ (i.e. the measurement results or the necessary operated qubits in all ports) to $E_1$. Now the remainder system is $R_0^3$ and its state is denoted as $\overline{[\rho^1\rho^2](n_1^2)}$, where the overline represents the system has been operated. Note that the qubits which was measured in $G^2$ are still in $R_0^3$, and the qubits which was sent to the other adversary (i.e. included in $L^2$) are replaced by the ones in certain predetermined states (e.g. $|0\rangle$ for every qubit). So the dimension and the port structure of $\overline{[\rho^1\rho^2](n_1^2)}$ are the same as that of $[\rho^1\rho^2](n_1^2)$.
\item[$A_2$] When $Q_1^2$ arrives, $E_1$ performs a measurement on the systems $R_1^2Q_1^2B_1^2$ to teleport $R_1^2Q_1^2$ which is in state $\rho^1(n_0^1)\rho^2$ with the outcome $n_1^2$. And then $E_0$ holds
    \begin{equation}\label{e3}
        [\rho^1(n_0^1)\rho^2](n_1^2)
    \end{equation}
    in $B_0^2$.
    As in $A_1$, suppose $E_0$ has performed the next teleportation (which should be really done when $Q_0^3$ arrives $E_0$ in Step $A_3$) and $E_1$ holds
    \begin{equation}\label{e4}
        \textbf{[}[\overline{\rho^1\rho^2](n_1^2)}\rho^3\textbf{]}(n_0^3)
    \end{equation}
    in $B_1^3$.
    For each port in $B_1^3$, $E_1$ only reserves the $n_1^2$-th sub-port and performs $G^3$ on it. Afterwards, $E_1$ sends the measurement results and the necessary operated qubits $L^3$ to $E_0$. Now the remainder system is $R_1^4$ and its state is denoted as $\overline{[\rho^1\rho^2\rho^3](n_0^3)}$.

\item[$A_i(3\leq i\leq n)$] $E_j$ performs a measurement on the systems $R_j^iQ_j^iB_j^i$ to execute a PBT of $R_j^iQ_j^i$ with a outcome $n_j^i$. And then $E_{j^\prime}$ holds
    \begin{equation}\label{e5}
        \textbf{[}\overline{[\rho^0\rho^1\ldots \rho^{i-1}](n_{j^\prime}^{i-1})}\rho^i\textbf{]}(n_j^i)
    \end{equation}
    in system $B_{j^\prime}^i$, where $j^\prime=j\oplus 1$.
  Suppose that $E_{j^\prime}$ has finished the next teleportation and $E_j$ holds
  \begin{equation}\label{e5}
    \textbf{[}\overline{[\rho^0\rho^1\ldots \rho^i](n_{j}^{i})}\rho^{i+1}\textbf{]}(n_{j^\prime}^{i+1})
  \end{equation}
  in $B_j^{i+1}$.
   For each port in $B_{j}^{i+1}$, $E_j$ first discards everything except the $n_j^i$-th sub-port to do the correction to the chaos introduced by herself in the teleportation she did above, then performs the operation $G^{i+1}$ on all the ports respectively. Afterwards, $E_j$ sends the measurement results and the necessary operated qubits $L^{i+1}$ to $E_{j^\prime}$. Now the remainder system is $R_j^{i+2}$ and its state is denoted as $\overline{[\rho^1\rho^2\ldots\rho^{i+1}](n_{j^\prime}^{i+1})}$.
\end{itemize}
\hspace{0.5cm}

In this attack, the adversaries send the classical signal from the verifiers and the which-port-information in their respective PBT (i.e. $n_j^i$) to each other once they were obtained. At the same time, the adversaries also send the response to their closer verifiers respectively once they recognized what is the correct response. It is not difficult to see that the correct and timely. Concretely, suppose that $Q_j^i$ will reach $P$ at time $t_i$, the distance between $E_0$ ($E_1$) and $P$ is $L$, and $\Delta T=L/c$ ($c$ is the speed of light). When $Q_1^2$ reaches $E_1$ at time $t_2-\Delta T$, the which-port-information $n_1^2$ is generated and sent to $E_0$, which will reach $E_0$ at $t_2+\Delta T$. Before this, at $t_1-\Delta T$, the possible results $L^2$ are generated and sent to $E_1$, which will reach $E_1$ at $t_1+\Delta T$. Once $E_0$ ($E_1$) gains both $n_1^2$ and $L^2$, she can deduce the correct response information. According to the arriving times of $n_1^2$ at $E_0$ and $L^2$ at $E_1$ (both $\leq t_2+\Delta T$), the responses are also in time. And this situation also holds for any later signal $Q_j^i$. Consequently, the attack strategy is successful.

Similar to that in the INQC attack in Sec.2.3, the above description of the attack is a general version, where some transmissions (e.g. some classical signals and $n_j^i$) or teleportations (e.g. the anterior qubits in $R_j^i$ which do not take any effect to the present operation) might be unnecessary when particular protocol is considered.

\subsection{The private-order DTQPV protocols}

In the above subsection, we proved the public-order DTQPV protocols are insecure. Consequently, QPV protocols has to be designed in the private-order model to pursue higher security. It is not difficult to see that the attack in the last subsection (i.e. Attack I) will fail if the order of the signals is unknown to $E_0$ and $E_1$ because the adversaries will not know when they need add an empty signal so that all the signals come to $P$ one-side by one-side [just like that in Eq. (\ref{e0})].

In fact, the private-order DTQPV protocol can be designed in different ways. For example, the manner of adding decoy signals, as discussed above, is one kind of private-order DTQPV protocols. Note that in such protocols $E_0$ and $E_1$ never know which side the next decoy signal will come from, which implies that the order of all signals is unknown to them. Here we give another kind of private-order DTQPV protocols, where the matching information is encoded into the chronological order of the challenge signals' arrivals at $P$ and no decoy signals are added. This kind of protocols are easier to analyze, and the conclusions about their security can be generalize to general DTQPV protocols. One example (called Protocol II) of these protocols is as follows.

\begin{itemize}
  \item[II1] In advance, $V_0$ and $V_1$ prepare a list of states $|\varphi_k\rangle_{ij}$ under an orthogonal basis, where $i,j=0,1$, $k=1,2,\ldots m$ and $\langle\varphi_k|\varphi_{k^\prime}\rangle=\delta_{kk^\prime}$ (i.e. mutually orthogonal with each other). Each state is divided into two subsystems, which will be sent to $P$ so that $P$ can measure them in the corresponding basis and then announce the result to the verifiers.
      If the subscripts $ij=00$ ($ij=11$), both subsystems in this state are stored in $V_0$'s ($V_1$'s) position. And if $ij=01$ or $ij=10$, each of the verifiers holds one of the two subsystems. What's more, the states are designed in such a way that each subsystem are in the same mixed state individually, just like the EPR pairs, thus only collective measurements can extract the information encoded in them correctly.\vspace{-2mm}
  \item[II2] When the protocol begins, $V_0$ and $V_1$ send every subsystems (one by one) to $P$ at random times but in such a way that the two subsystems of the same state arriving at $P$  adjacently, i.e. no other subsystems arrive at $P$ between them.\vspace{-2mm}
  \item[II3]Whenever $P$ receives two new subsystems, he measures them in the bases $\{|\varphi_1\rangle,|\varphi_2\rangle,\ldots,|\varphi_m\rangle\}$ and sends the measurement result to both $V_0$ and $V_1$ immediately.\vspace{-2mm}
  \item[II4] $V_0$ and $V_1$ jointly accept $P$'s position if and only if they received all of $P$'s response results correctly and in time.
\end{itemize}

Since the order of the challenge signals in Protocol II are unknown to the adversaries, they cannot know how to insert the ``empty signals''. Consequently, they cannot normalize the challenge signals in the form of Eq. (\ref{e0}), which will cause the situation that each adversary is unaware of the order of the signal which she is teleporting in the whole signal sequence. And through a concrete analysis, we find if the adversaries follows the steps $A_1$ to $A_n$ in Attack I to attack Protocol II, but without a chronological order of all the signals as in Eq. (\ref{e0}), they cannot make the response information reach $V_0$ and $V_1$ in time.

\emph{In fact, Protocol II is secure against all the attack model where the adversaries perform the attack operations only when they detect the arrival of the challenge signals.}
To prove this, we consider the first two signals sent by $E_0$, denoted as $Q_0^1$ and $Q_0^2$, and the first two sent by $E_1$, denoted as $Q_1^1$ and $Q_1^2$.
There are three possible operations that $P$ will perform: measuring $Q_0^1Q_0^2$, measuring $Q_0^1Q_1^1$, and measuring $Q_1^1Q_1^2$. The correct choice is encoded by the time of these signals' arrivals at $P$. If the adversaries want to measure them correctly, they must compare their arrival time.
However, the challenge signal's arrival time can only be known to the adversaries when it indeed reaches $E_0$ or $E_1$. To consider the four signals as a whole, $E_0$ ($E_1$) can deduce the correct order information no sooner than $E_1$'s ($E_0$'s) classical information for teleporting $Q_1^2$ ($Q_0^2$) arrives at $E_0$ ($E_1$). And only then, they can deduce the correct response information. So due to the limitation of  the no signaling theorem, the responses will be late doubtlessly. The similar problems will also hold in the following processes. So Protocol II is secure against the adversaries who perform the attack operations only when they detect the arrival of the challenge signals. Considering the strategy of sending signals at random time is employed, we get the conclusion that Protocol II is secure against INQC attack for the \emph{general} adversaries, who execute the attack operations at every common time slot or the time when they detect the arrival of the challenge signals from the verifiers, just like Protocol I.

As an instance of private-order QPV, Protocol II has been proved secure against the general adversaries.
However, facing the same problem with Protocol I, this type of QPV is also insecure against the SINQC attack. Actually, through adding the ``empty signals'' with extremely high frequency during the whole process of the protocol, the adversaries with strong ability can still list the challenge signals in the form of Eq. (\ref{e0}). That is to say, through the strategy above, adversaries can transform any private-order QPV protocol into a public-order one and then attack it as a public-order one successfully utilizing (the superdense version of) Attack I. And we denote this improved vision of Attack I as Attack II. Obviously, as private-order QPV protocols, both Protocol II and the protocols utilizing decoy signals are insecure against Attack II.

In a word, to analyze the situation of breaking down H3 and H4, we find an interesting fact that \emph{there exist normal-qubit-based QPV protocols with the same security level of (wave-packet-based) Protocol I, that is, the adversaries have to perform SINQC to attack successfully}. Considering the difficulty of realizing the single-photon interference for long distance, private-order DTQPV protocols, such as Protocol II and the protocols utilizing decoy signals, are more practical.

\section{Protocol III: Attempt to break down all the four hidden assumptions}

\begin{figure}
  \center
  \includegraphics[width=14cm]{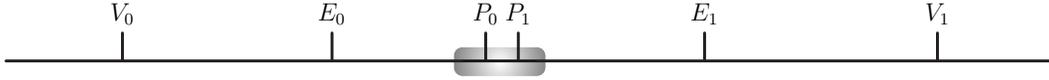}\\
  \caption{Protocol III: the two-point QPV employing single-photon interference and random sending time.}\label{figIII}
\end{figure}

Now we will continue to consider the situation of breaking all of the hidden assumptions. To make the matching information unclear to the adversaries in the QPV protocol employing single-photon interference, we can set up more than one interference points in the secure area of $P$. Let us take the two-point one as an example (See Fig. \ref{figIII}), where the two interference points are labeled as $P_0$ and $P_1$. The wave packets on the two sides may simultaneously arrive at either $P_0$ or $P_1$. During the protocol which interference point works in the current round is determined by the measurement results of the previous rounds. For example, if the measurement result of the last round is 0 (1), $P_0$ ($P_1$) will work in the current round. Thus, the matching information is uncharted for the adversaries and the attack strategy to Protocol I will fail here.

We call the QPV protocols with two or more proving points as multi-point QPV. In this kind of protocols, there is one receiving device at every proving point, and all of them are controlled (on or off) by the prover $P$. Generally the distance between different proving points is very short and the protocol can be easily realized. This modification does not damage the security of the protocol when $P_0$ and $P_1$ are both in the secure region around $P$. Note that there are difference between the multi-point strategy and the private-order DTQPV. The challenge signals are never delayed (i.e. stored by $P$ waiting its matched signal) in the former while some of the challenge signals have to be delayed in the latter, which means the prover should store a part of the signals before operations on them.
Consequently, Attack II is likely to be ineffective for the multi-point protocols. In fact, if the adversaries employ Attack II to attack the multi-point QPV protocols, by a detailed analysis we find that the response information will be delayed definitely when the interference point is supposed to be changed.

Obviously, the multi-point QPV with single-photon interference and random sending time is an improved version of Protocols I and II, so it might be more difficult to attack. However, it is still insecure against the special attack strategy (i.e. SINQC) in theory though the attack becomes more complex. Next, we will take the two-point one (denoted as Protocol III) as our example to present the particular attack strategy (denoted as Attack III) in detail.

Without loss of generality, suppose the scheme begin with a pair of matched challenge signals (i.e. the packets divided from a photon) which are sent by $V_0$ and $V_1$ so that they will arrive at $P_1$ simultaneously, and the time when $E_0$ and $E_1$ received their first matched challenge signals is $t_0$ (suppose the positions of $E_0$ and $E_1$ are symmetrical with respect to $P_1$).
For the sake of clarity, we list the notations that we will use in the following description in Table \ref{tab3}. And for simplicity, the following description omits the transmission of classical information. In fact, the adversaries send all the necessary classical information to each other, including the which-port-information generated in the teleportation and the Bell-measurement results (i.e. the interference results) on the teleported systems. The detailed process of Attack III is as follows.
\begin{table}
\centering
\caption{The notations in the description of Attack III. Here, $t_i=t_0+i\Delta t$ and $j=0,1$.}
\label{tab3}\vspace{1mm}
\renewcommand{\arraystretch}{1.15}
\begin{tabular}{c|l}
  Notation & What the notation denotes \\
  \hline\hline
  \multirow{1}*{$L$} & The distance between $P_0$ and $P_1$.\\\hline
  \multirow{2}*{$\Delta t$} &  The maximum time interval for the adversaries to perform attacks to ensure no wave\\
                         & packets are missed (i.e. SINQC attack), which is ultra short. \\\hline
  \multirow{2}*{$Q_j^{t_i}$} & The quantum signal $E_j$ receives at $t_i$ (suppose it has been transformed into normal qubit\\
                             & from the wave packet \cite{Aharonov}). For simplicity, it also means the state of this signal.\\\hline
  \multirow{3}*{$C_j^{t_i}$} & The Bell-measurement results obtained in the previous rounds on $E_j$'s side at $t_i$ (actually\\
                         & the part which has been received by the counterpart can be excluded). For simplicity, it also\\
                         & means the quantum state carrying these results.\\\hline
  \multirow{2}*{$B^{t_i}$} & The pre-shared entangled systems $E_0$ uses at $t_i$ ($E_1$ will use a part of them before $t_i$). Note\\
                           & that the entangled states are enough and every PBT via them consumes different part of them.\\\hline
  \multirow{1}*{$n_j^{t_i}$} & The which-port-information of the teleportation $E_j$ performed at $t_i$.\\\hline
  \multirow{2}*{$Q(n_j^{t_i})$} & The resulted multi-port system after the counterpart's measurement in the PBT. The\\
                             & dimensionality of every port is equal to that of $Q$ and the state of the $n_j^{t_i}$-th port is just $Q$.\\\hline
\end{tabular}
\end{table}

\begin{itemize}
  \item[III1] At $t_0$, $E_0$ teleports the received signal $Q_0^{t_0}$ to $E_1$ through the entangled system $B^{t_0}$ with the outcome $n_0^{t_0}$. Thus $E_1$ obtains $Q_0^{t_0}(n_0^{t_0})$. Then $E_1$ teleports it back (via the entanglement in $B^{t_0}$) to $E_0$ together with the signal she received $Q_1^{t_0}$, with the outcome $n_1^{t_0}$. Then $E_0$ clears away the redundant systems according to $n_0^{t_0}$ and obtains $[Q_0^{t_0},Q_1^{t_0}](n_1^{t_0})$. $E_0$ measures each port of it and gets the measurement results.\vspace{-2mm}
  \item[III2] For the period from $t_0$ to $t_L=t_0+2L/c$, the adversaries perform the following operations at regular intervals $\Delta t$, i.e. at every $t_i=t_0+i\Delta t$, where $i=0,1,...,\frac{2L}{c\Delta t}$.

      \textbf{(III2.1)} $E_0$ teleports $Q_0^{t_i}$ to $E_1$ through $B^{t_i}$ and then the latter gets $Q_0^{t_i}(n_0^{t_i})$. (If no wave packet arrives, $Q_0^{t_i}$ is in state $|0\rangle$, which is the original state of the normal qubit when we want to transform a wave-packet state into a normal-qubit one \cite{Aharonov}.)

      \textbf{(III2.2)} $E_1$ teleports $Q_0^{t_i}(n_0^{t_i})$ back to $E_0$ (via the entangled states in $B^{t_i}$) together with $Q_1^{t_i}$ and the classical information obtained in the previous rounds $C_1^{t_i}$. $E_0$ obtains $[Q_0^{t_i}(n_0^{t_i}),Q_1^{t_i},C_1^{t_i}](n_1^{t_i})$ and then clears away $n_0^{t_i}$ and obtains $[Q_0^{t_i},Q_1^{t_i},C_1^{t_i}](n_1^{t_i})$.

      \textbf{(III2.3)} For each port, $E_0$ first measures the classical information $C_1^{t_i}$. Combining with her own $C_0^{t_i}$, $E_0$ can deduce whether the two signals are matched. For the ports in which the signals are matched, $E_0$ measures them in Bell bases; and for the ports in which the signals are not matched, she retains them.

     \textbf{Note}. The unmatched situation is relatively intricate. It implies the wave packets will reach $P_0$ simultaneously. So no wave packet arrives at $E_0$ currently (otherwise $E_1$ should receive its matched packet before $t_0$, which contradicts with the postulate that the protocol begins at $t_0$). Consequently, $Q_1^{t_i}$ should be matched with $Q_0^{t_{i+}}$, where $t_{i+}=t_i+2L/c$. Generally, they should wait until $E_0$ receive $Q_0^{t_{i+}}$ and measure $Q_0^{t_{i+}}$ and $Q_1^{t_i}$ in Bell bases. The concrete operations are: (a) $E_0$ teleports $Q_0^{t_{i+}}$, $C_0^{t_i}$ and $Q_1^{t_i}(n_1^{t_i})$ to $E_1$ through $B^{t_{i+}}$; (b) $E_1$ clears away $n_1^{t_i}$ and gets $[Q_0^{t_{i+}},Q_1^{t_i},C_0^{t_i}](n_0^{t_{i+}})$ at the corresponding system of $B^{t_{i+}}$; (c) for each port, $E_1$ first measures $C_0^{t_i}$ and gets the matching information combining with $C_1^{t_i}$, and then she measures $Q_0^{t_{i+}}$ and $Q_1^{t_i}$ in the unmatched port in Bell basis (the matched port will be overlooked now, which implies $Q_0^{t_{i}}$ and $Q_1^{t_i}$ are matched and was already measured by $E_0$ at $t_i$). However, if the adversaries execute the above operations until $t_{i+}$, the response to $V_0$ would be delayed by $2L/c$. How to avoid this? As we know, the probability distributions of the measurement results is independent with the measurement order. Consequently, operation (a) has to be executed until $t^{i+}$, while (b) and (c) do not. So $E_1$ can execute (b) and (c) at $t_i$ before (a) has finished.

      \textbf{(III2.4)} $E_1$ images that she has obtained the state
      \begin{equation}
        [Q_0^{t_{i+}},Q_1^{t_i}(n_1^{t_i}),C_0^{t_i}](n_0^{t_{i+}}),
      \end{equation}
      which is teleported from $E_0$, in $B^{t_{i+}}$, and does the corrections and measurements to the system as described in (b) and (c) above.\vspace{-2mm}
  \item[III3] For the period after $t_L$ (until the end), the adversaries perform the following operations at every $t_i=t_L+k\Delta t$, where $i=(2L)/(c\Delta t)+k$ and $k>0$.

  \textbf{(III3.1)} $E_0$ teleports the signal currently received by her $Q_0^{t_i}$, the reserved $Q_1^{t_{i-}}(n_1^{t_{i-}})$ at $t_{i-}=t_i-2L/c$, and the previous classical information $C_0^{t_{i-}}$ to $E_1$ through $B^{t_i}$. This is coincident with the operation (a) above.

  \textbf{(III3.2)}-\textbf{(III3.4)} are the same as \textbf{(III2.2)}-\textbf{(III2.4)}, respectively.
\end{itemize}

In this attack, the adversaries send the response to their closer verifiers respectively once they recognized what is the correct response. Similar to Attack I, the adversaries' response will be correct and timely, and so this attack will be successful.

Here Protocol III, i.e. the two-point QPV protocol employing single-photon interference and random sending time, can be modified to multi-point ones. However, Attack III can also be easily extended to the corresponding versions which can successfully attack such protocols.
And it is worth noting that this strategy, i.e. the multi-point strategy, can also be used to modify the private-order DTQPV protocols and such protocols would still be insecure against the corresponding attacks which would be the combinations of Attack II and Attack III. We will not describe such protocols and attacks in detail since they are just simple extensions and combinations.

\section{Conclusions}
In this paper, we attempted to overcome a crucial obstacle in the study of pursuing secure QPV, i.e. the no-go theorem proposed by Buhrman et al. \cite{Buhrman}. By analyzing four implicit and essential hidden assumptions for a successful INQC attack, we proposed three kinds of protocols which are not consistent with these assumptions.
Our results show that for the \emph{general} adversaries, who execute the attack operations at every common time slot or the time when they detect the arrival of the challenge signals from the verifiers, our proposed protocols are secure even if unlimited pre-shared entanglement is allowed to the verifiers (i.e. the standard secure model).

However, as we demonstrated, these protocols are still insecure in theory because they are vulnerable to the SINQC attack, where the adversaries have to launch a new round of attack on the coming qubits with extremely high frequency so that none of the possible qubits, which was sent at random times, will be missed.
In spite of this, our proposed protocols still achieve \emph{practically} security because of the high difficulty of realizing the SINQC attack. Furthermore, since most of the entanglement resource pre-shared by the adversaries are wasted to teleport the ``empty signals'', including the state $|0\rangle$ when no wave packet arrives in the protocols with single-photon interference (e.g, protocols I and III) and the added empty signals in private-order DTQPV (e.g. protocol II), the adversaries need much more entanglement to execute a successful attack to our protocols than the previous ones. It implies our protocols also exhibits higher security when the amount of the shared entanglement among the adversaries is bounded (i.e. the bounded secure model).

By proposing the three kinds of new QPV protocols and the corresponding SINQC attack strategies, we have actually proven the impossibility of secure QPV with looser assumptions, which means the breach of the four hidden assumptions H1-H4. On this basis, we propose the enhanced no-go theorem for QPV.

\textbf{The enhanced no-go theorem for QPV.}
Besides the previous ones \cite{KMS10,CFG10,M101,LL11}, QPV protocols employing the following strategies are still insecure in theory, that is, all of them are still vulnerable to the SINQC attack. (1) The strategy of employing wave packets of the single-photon interference and random sending time, such as Protocol I; (2) The strategy that the arrival times at $P$ of the matched signals are different, such as the private-order DTQPV (i.e. protocol II); (3) The strategy of setting up more than one proving points where the matched signals should arrive simultaneously, i.e. the multi-point QPV; (4) The combining of the above three strategies, such as the protocols designed by combining the breach of H1 and H3 (Protocol III), and the protocols designed by combining the breach of H2 and H3.

Note that in this paper we used one dimension QPV as our example to demonstrate the protocols and the attacks. In fact the conclusions can also be generalized into the multi-dimension QPV protocols.

At last, we would like to list some other conclusions about QPV.\vspace{-2mm}
\begin{itemize}
  \item The results of local measurements to different systems are independent with the order of the measurements. This property plays an crucial role in designing our proposed attack strategies and should be paid attention to in the analysis of QPV.\vspace{-2mm}
  \item By using single-photon interference and random sending time (i.e. breaking down H1 and H2), the adversaries in INQC attack is forced to be blind. That is to say, they have to employ SINQC to perform a successful attack, and the difficulty for SINQC is just like that a blind man, who stands beside a fiber where photons will pass by at random time, tries to ``cut'' the fiber by a knife continually so that he can ``catch'' all the possible photons. And this is a new level of security for QPV. \vspace{-2mm}
  \item It is not difficult to understand that the adversaries have to perform SINQC to attack the protocols with single-photon interference (e.g. Protocol I) since the adversaries cannot be aware of the arrival of the signal (otherwise the interference would be damaged). But it is interesting that protocols employing normal qubits can also reach this level of security (that is, can only be successfully attacked by SINQC), such as the private-order DTQPV, which does not add any additional difficulty in the realization compared with the previous QPV protocols).\vspace{-2mm}
\end{itemize}

\section*{Acknowledgements}
We are grateful to Christian Schaffner, Serge Fehr, Hoi-Kwan Lau, Hoi-Kwong Lo, Gilles Brassard, Xiongfeng Ma, Qingyu Cai, Zhangqi Yin, Changling Zou for useful discussions. This work is supported by NSFC (Grant Nos. 61272057, 61202434, 61170270, 61100203, 61003286, 61121061), NCET (Grant No. NCET-10-0260), Beijing Natural Science Foundation (Grant Nos. 4112040, 4122054), the Fundamental Research Funds for the Central Universities (Grant No. 2012RC0612, 2011YB01), BUPT Excellent Ph.D. Students Foundation (Grant CX201216).



\begin{thebibliography}{0}
\bibitem{PBC} Chandran, N., Goyal, V., Moriarty, R., Ostrovsky, R.: Position based cryptography. In: Halevi, S. (Ed.) CRYPTO 2009. LNCS, vol. 5677, pp. 391¨C407. Springer, Heidelberg (2009)
\bibitem{Pat} Kent, A., Munro, W., Spiller, T., Beausoleil, R.: Tagging systems, US patent nr 2006/0022832 (2006)
\bibitem{KMS10} Kent, A., Munro, B., Spiller, T.: Quantum tagging: Authenticating location via quantum information and relativistic signalling constraints. Phys. Rev. A 84, 012326 (2011), also at arXiv/quant-ph:1008.2147 (2010)
\bibitem{CFG10} Chandran, N., Fehr, S., Gelles, R., Goyal, V., Ostrovsky, R.: Position-based quantum cryptography, arXiv/quant-ph:1005.1750 (2010)
\bibitem{M101} Malaney, R.A.: Location-dependent communications using quantum entanglement. Phys. Rev. A 81, 042319 (2010)
\bibitem{LL11} Lau, H.K., Lo, H.K.: Insecurity of position-based quantum-cryptography protocols against entanglement attacks. Phys. Rev. A 83, 012322 (2011)
\bibitem{Buhrman} Buhrman H., Chandran N., Fehr S., Gelles R., Goyal V., Ostrovsky R., and Schaffner C.: Position-Based Quantum Cryptography: Impossibility and Constructions. In: Rogaway P. (Ed.) CRYPTO 2011. LNCS, vol. 6841, pp. 429¨C446. Springer, Heidelberg (2011)
\bibitem{Nature} Brassard G.: The conundrum of secure positioning. Nature 479, 307 (2011)
\bibitem{BK11} Beigi S., and Konig R.: Simplified instantaneous non-local quantum computation with applications to position-based cryptography. New J. Phys. 13, 093036 (2011)
\bibitem{gardenhose} Buhrman H., Fehr S., Schaffner C., and Speelman F.: The Garden-Hose Model. arXiv/quant-ph:1109.2563 (2011)
\bibitem{Kent11} Kent A.: Quantum tagging for tags containing secret classical data. Phys. Rev. A 84, 022335 (2011)
\bibitem{GV95} Goldenberg L., and Vaidman L.: Quantum cryptography based on orthogonal states. Phys. Rev. Lett. 75, 1239 (1995)
\bibitem{Complem} Busch P., and Shilladay C.R.: Complementarity and Uncertainty in Mach-Zehnder Interferometry and beyond. Phys. Rep. 435, 1 (2006)
\bibitem{PBT1} Ishizaka S., and Hiroshima T.: Asymptotic teleportation scheme as a universal programmable quantum processor. Phys. Rev. Lett. 101, 240501 (2008)
\bibitem{PBT2} Ishizaka S., and Hiroshima T.: Quantum teleportation scheme by selecting one of multiple output ports. Phys. Rev. A 79, 042306 (2009)

\bibitem{Aharonov} Aharonov Y., and Vaidman L.: Nonlocal aspects of a quantum wave. Phys. Rev. A 61, 052108 (2000)

\end{thebibliography}
\end{document}